# Uncovering A Two-Dimensional Semiconductor with Intrinsic Ferromagnetism at Room Temperature

Fang Zhang[1,2], Xing-Qiang Shi[2, *], Zi-Kang Tang[1, *]

[1]Institute of Applied Physics and Materials Engineering, University of Macau, Macau, China
[2]Department of Physics, Southern University of Science and Technology, Shenzhen, China.

**\*Corresponding Authors**
   shixq@sustech.edu.cn (X. Q. Shi)
   zktang@um.edu.mo (Z. K. Tang)

**Two-dimensional materials have been gaining great attention as they displayed a broad series of electronic properties that ranging from superconductivity to topology. Among them, those which possess magnetism are most desirable, enabling us to manipulate charge and spin simultaneously. Here, based on first-principles calculation, we demonstrate monolayer chromium iodide arsenide (CrIAs), an undiscovered stable two-dimensional material, is an intrinsic** 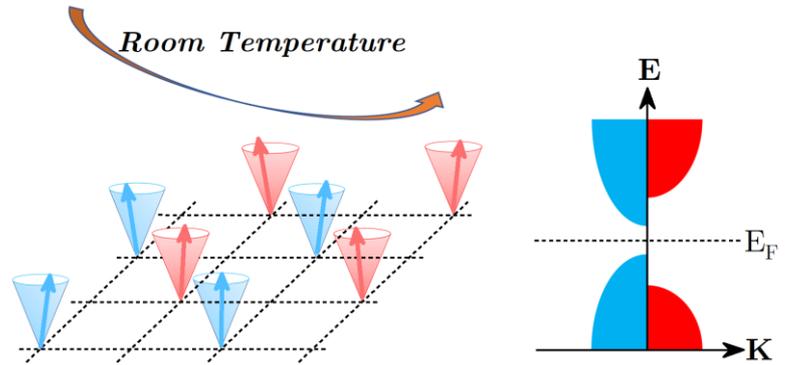 **ferromagnetic semiconductor with out-of-plane spin magnetization. The indirect bandgaps are predicted to be 0.32 eV for majority spin and 3.31 eV for minority spin, large enough to preserve semiconducting features at room temperature. Its Curie temperature, estimated by Heisenberg model with magnetic anisotropic energy using Monte Carlo method, is as high as 655 K that well above the room temperature, owing to strong direct exchange interaction between chromium *d* and iodine *p* orbitals. This work offers the affirmative answer of whether there exists two-dimensional ferromagnetic semiconductor at room temperature. And the practical realization of quantum spintronic devices, which have been suppressed because of lacking suitable room temperature magnetic materials, would embrace a great opportunity.**

***Introduction*** Since the discovery of graphene[1], the upsurge of two-dimensional (2D) materials has never been ceased. Novel phenomena continue to emerge, including superconductors[2-4], Mott insulators[5] and topological semimetals[6]. Within 2D materials, ferromagnetic (FM) semiconductors, of which integrate ferromagnetism and semiconductivity, are most desirable. They provide us the chance to manipulate electron's charge and spin simultaneously in confined dimensions, which serves as prerequisite for realizing quantum spintronic devices[7-8]. Moreover, the heterostructures that incorporate topological material with FM semiconductor, are a new proposed structure to observe quantum anomalous Hall effect, thus making them the prototype of dissipationless devices[9]. Regrettably, not even one 2D material coalescing all features up mentioned has been discovered at room-temperature[10], despite of great effort scientist have putted into.

In the early stage, 2D FM semiconductors were believed not exist. Two-dimensionality and FM order in isotropic Heisenberg model at finite temperature are mutually incompatible, stood by Mermin-Wagner theorem[11]. The emergence of monolayer CrI$_3$[12] and bilayer Cr$_2$Ge$_2$Te$_6$[13] have proven that a slight magnetocrystalline anisotropy is enough to remove the restriction the theorem sets. Although the ferromagnetism of them are only maintained down to low temperature due to weak super-exchange interaction, the possibility of intrinsic high temperature 2D magnets are not fundamentally excluded[14]. Preservation of ferromagnetism at room temperature in monolayer metallic state VSe$_2$[15], as well as 2D semiconducting state Fe$_3$GeTe$_2$ obtaining by gate control[16], are indubitable evidences to suggest the existence of 2D intrinsic room-temperature FM semiconductors.

Several structures are recently predicted to be room temperature semiconducting ferromagnets, such as antiaromatic ring based organometallic framework[17] and PdX$_3$[18] (X=Br, Cl). However, they suffer from drawbacks. The former one requires magnetic transition metals connected by organic linkers, which is difficult to make devices because of its complex structure and stability. The latter is a single 2D crystal, but it remains highly possible that the crystal has metallicity instead of semiconductivity at room temperature, considering its extremely narrow bandgap. Sr$_2$Fe$_{1+x}$Re$_{1-x}$O$_6$[19] is a newly synthesized room temperature ferromagnet, but double-perovskite crystal limits it down to 2D structure. Therefore, intrinsic 2D FM semiconductor at room temperature is still unrevealed.

Here, we show that, by state-of-art first-principles calculation and Monte Carlo simulation, monolayer CrIAs is a stable intrinsic 2D FM semiconductor at room temperature. Monolayer CrIAs is dynamical and thermal stable, with bandgap ($E_g$) 0.32 eV for majority spin, 3.31 eV for minority spin, and Curie temperature ($T_c$) of 655 K. $E_g/k_B$ and $T_c$ are large enough to preserve semiconducting and FM features at room temperature. We concluded that the large $T_c$, which is mainly dominated by first nearest exchange interaction between metal ions, are originated from extremely strong direct exchange between Cr $d$ orbitals and I $p$ orbitals. The discovery of CrIAs is expected to stimulate researchers from 2D magnetic area, and to make 2D magnetic semiconductors toward room temperature application.

***Methods*** Our spin-polarized first-principles calculations were carried out by using the projector-augmented-wave (PAW) potential[20] and the generalized gradient approximation (GGA) in the form proposed by Perdew, Burke, and Ernzerhof (PBE)[21], as implemented in Vienna *ab initio* simulation package (VASP)[22]. A sufficiently large vacuum space of 15 Å was used. All calculations employed the cut-off energy of 600 eV, convergence criteria for the energy difference in electronic self-consistent loop 10$^{-6}$ eV and residual forces on ions less than 10$^{-3}$ eV/Å. A Monkhorst-Pack $k$-point mesh with $\Gamma$-centered of 6×9×1 was used for primitive cell. Since the GGA cannot properly describe strongly correlated systems containing partially filled $d$ or $f$ subshells (chromium here), and systemic properties are strongly related to U value which is hard to decide in GGA+U method[23-25], we employed the screened hybrid Heyd, Scuseria,

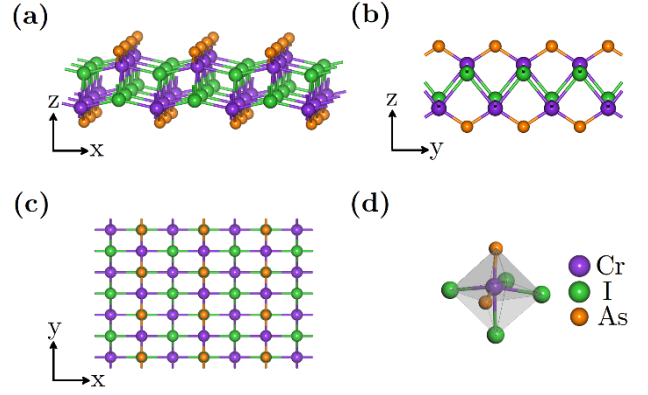

**Figure 1. Atomic structure of monolayer CrIAs.** (a) Front view, (b) side view and (c) top view of monolayer CrIAs displayed in 3 × 3 × 1 supercell. (d) Distorted octahedral structure of Cr atom.

Ernzerhof (HSE06) functional[26-27] that includes the accurate Hartree-Fock exchange energy despite of huge computational cost. Phonon band structure and DOS picture are calculated with density functional perturbation theory (DFPT)[28] by the Phonopy package interfaced to VASP code with 3×3×1 supercell. To estimate the Curie temperature by Heisenberg model with anisotropic energy, Monte Carlo simulation[29-30] was conducted with 64×64×1 supercell. For each temperature scale, 2×10$^7$ steps were taken to get thermodynamic equilibrium as well as a reasonable average value. More calculation details are addressed in *supporting information* (*SI*) section (Sec.) 1.

***Structure and Stability*** The structure of monolayer CrIAs is depicted in Figure (Fig.) 1, whose space group is Pmmn (No. 59). Lattice constants are $a$ = 5.321 Å, $b$ = 3.717 Å, $\alpha = \beta = \gamma$ = 90°, belongs to orthorhombic system (method of obtaining lattice constants is in Sec. 2, *SI*). Each primitive cell contains two formula units of CrIAs and has two Cr atoms. Since the crystal filed of the surrounding I and As is distorted octahedra, it is inappropriate to split the *d* orbitals of the two Cr atoms into threefold $t_{2g}$ and twofold $e_g$ orbitals. However, splitting *d* orbitals to $d_{xy}$, $d_{xz}$, $d_{yz}$, $d_{z^2}$ and $d_{x^2+y^2}$ is still valid, which is guaranteed by quantum mechanics[31-32]. For $Cr^{4+}$ with two electrons in monolayer, two electrons fill different and low-energy *d* orbitals, as Hund's law requires maximum spin moment and Pauli Exclusion Principle forbids electrons with same spin occupying the same orbital. Therefore, the spin per metal atom is **S** = 2/2. To confirm the stability of monolayer CrIAs, its phonon spectra have been calculated. There is no imaginary frequency in the whole Brillouin zone (Fig.2), indicating CrIAs monolayer is kinetically stable. Moreover, molecular dynamics have been performed with canonical ensemble (Sec. 4 in *SI*) to evaluate the thermodynamic stability. Total energy and magnetic moments resist thermal fluctuation at room temperature, further indicating that monolayer CrIAs is experimentally synthesizable.

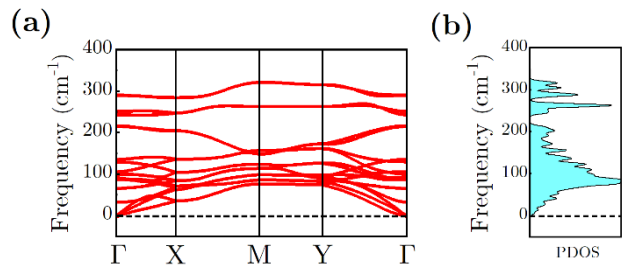

**Figure 2. Phonon spectra of monolayer CrIAs.** (a) Phonon band structure through first Brillouin zone high symmetry points. (b) Phonon density of states (PDOS) of 3×3×1 supercell. No imaginary parts indicating monolayer is kinetically stable.

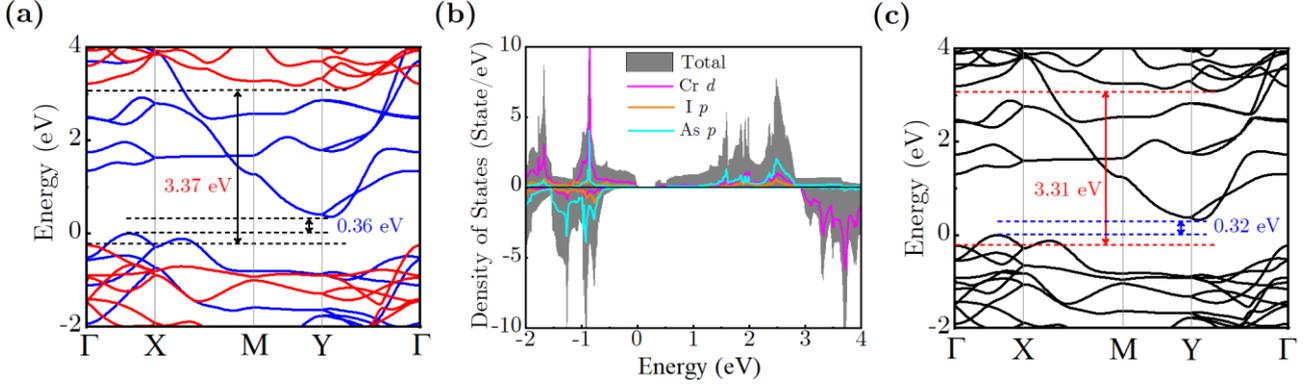

**Figure 3. Electronic properties of monolayer CrIAs at HSE06 level.** (a) Band structure of monolayer CrIAs of high symmetrical paths in first Brillouin zone, (b) density of states and (c) band structure with spin-orbit coupling (SOC). Monolayer CrIAs is a ferromagnetic semiconductor. The indirect bandgap of minority spin is 0.32 eV, of majority spin is 3.31 eV, both large enough to make monolayer CrIAs maintaining semiconducting properties at room temperature. All Fermi levels are set to be valence band maximum (VBM) at energy zero.

*Electronic Properties* To determine the ground magnetic state of CrIAs monolayer, we optimized the atomic position and calculated the total energy at HSE06 level for FM and antiferromagnetic (AFM) configurations, and found the FM state has an energy that lower than AFM state. The total magnetic moment per Cr site is +2.0$\mu_B$, while local magnetic moments for I and As are -0.13$\mu_B$ and -0.69$\mu_B$, respectively. The electronic band structure with the absence of spin-orbit coupling (SOC) (Fig. 3a) shows that CrIAs is a semiconductor with indirect bandgaps 0.36 eV for majority spin and 3.37 eV for minority spin. From the calculated atom projected density of states (Fig. 3b), the occupied state near Fermi level is mainly contributed by Cr *d* orbitals. Since I is a heavy element, the electronic structure with SOC (Fig. 3c) is also studied. Although bandgaps decrease 0.04 eV for majority spin and 0.06 eV for minority spin, they are still large enough to withstand thermal fluctuation at room temperature of 26 meV. The total magnetic moment per ion site increase 0.01$\mu_B$ when counting SOC. Compared to 2.0$\mu_B$, it is extremely small and thus neglectable.

*Magnetic Interaction* To gain a deeper level of understanding the magnetic properties of monolayer CrIAs, we adopt a Heisenberg spin Hamiltonian with magnetic anisotropies without external magnetic field that written as

$$H = -\frac{1}{2}\sum_{i,j} J_{ij} \boldsymbol{S_i} \cdot \boldsymbol{S_j} - \sum_i A(S_i^z)^2$$

in which $\boldsymbol{S_i}$ is the spin operator on site *i* and $|\boldsymbol{S_i}|=1$. $J_{ij}$ is the exchange constants between sites *i* and *j*, where positive value represents the FM coupling. *A* is the single-ion anisotropy with value *A* = 6.1 meV calculated at HSE06 level (Table S1). $\boldsymbol{S^z}$ denotes the spin orientation towards out-of-plane direction, namely *z*-axis. Factor ½ accounts for the double summation *i→j* and *j→i*.

To obtain the three nearest magnetic exchange constants, we designed four different magnetic orientations, including one FM state and three AFM states (denoted as $AFM_1$, $AFM_2$, $AFM_3$), as shown in Fig. S2, *SI*. The first three neighboring exchange constants $J_1$, $J_2$ and $J_3$ with interaction distance $d(J_1) < d(J_2) < d(J_3)$ could be obtained by matrix,

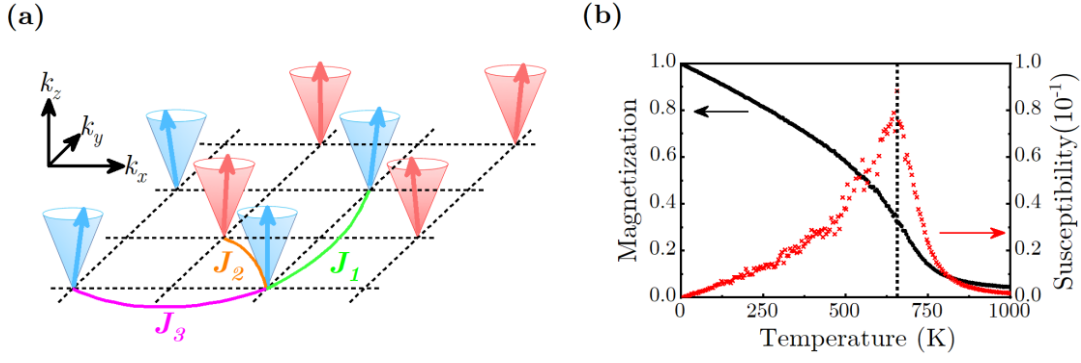

**Figure 4. Curie temperature of monolayer CrIAs.** (a) Diagrammatic illustration of magnetic exchange constants $J_1$, $J_2$ and $J_3$. Monolayer CrIAs has two Cr sublayers (see Fig.1a), the Cr sites in each sublayer are represented by red and blue cones. Blue and red arrows represent local spin direction. Based on the distance between Cr sites, magnetic interaction is divided into $J_1$, the nearest interaction; $J_2$, the second nearest; and $J_3$, the third nearest. Under different level thermal fluctuation, the spin direction would change accordingly. (b) The normalized magnetization (black) and magnetic susceptibility (red) with respect to temperature in the range of 0K to 1000K with step of 5K, obtained by Monte Carlo simulation using Heisenberg model in 64 × 64 × 1 supercell. The estimated Curie temperature is 655 K, much higher than room temperature.

**Table 1. Total energy per Cr site for monolayer with different magnetic configurations at HSE06 level.** The unit is meV. $E_{FM}$ is the reference energy and set to zero.

|  | $E_{FM}$ | $E_{AFM1}$ | $E_{AFM2}$ | $E_{AFM3}$ | $J_1$ | $J_2$ | $J_3$ |
|---|---|---|---|---|---|---|---|
| CrIAs | 0 | 5.2 | 581.8 | 37.7 | 289.6 | 1.3 | 17.6 |

$$\begin{pmatrix} J_1 \\ J_2 \\ J_3 \end{pmatrix} = \begin{pmatrix} -1/4 & -1/4 & 1/2 & 0 \\ -1/4 & 1/4 & 0 & 0 \\ -1/4 & -1/4 & 0 & 1/2 \end{pmatrix} \begin{pmatrix} E_{FM} \\ E_{AFM1} \\ E_{AFM2} \\ E_{AFM3} \end{pmatrix}$$

where $E_{FM}$, $E_{AFM1}$, $E_{AFM2}$ and $E_{AFM3}$ are the totally energies of magnetic orientations FM, AFM$_1$, AFM$_2$ and AFM$_3$, respectively. The derivation of above matrix is in Sec. 3, *SI*. In monolayer CrIAs, there has two identical Cr sublayers as shown in Fig. 1a and Fig. 4a. Each Cr site has eight neighboring interactions, two $J_1$, four $J_2$ and two $J_3$. The value of total energies and magnetic interactions are shown in Table 1.

To obtain $T_c$, we performed Monte Carlo simulation by using Heisenberg model. In Heisenberg model, spin is allowed to rotate randomly towards all directions, not bounded to $\pm z$ directions as Ising model restricts. And it is able to take magnetocrystalline anisotropy into account. Therefore, Heisenberg model is more general and valid for 2D materials[33]. The input variables including three magnetic exchange constants $J_1$, $J_2$, $J_3$ and magnetic anisotropy $A$. From the normalized magnetization and susceptibility (Fig. 4b) with respect to temperature, we extracted $T_c \approx$ 655K, which significantly exceeds the room temperature of 300K.

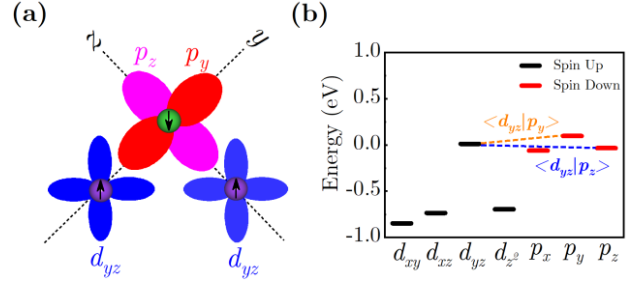

**Figure 5. Direct exchange mechanism.** (a) Diagrammatic illustration of Cr $d_{yz}$ and I $p_x$, $p_y$ orbitals towards $z$ axis. $d_{yz}$ orbital on the left could interact with $p_y$ orbital, while $d_{yz}$ orbital on the right could interact with $p_x$ orbital. (b) Average energy level of selected orbitals. Cr $d_{yz}$ orbital has similar energy level to $p_x$, $p_y$ orbitals.

***Direct Exchange mechanism*** Since the nearest magnetic interaction $J_1$ is much stronger than $J_2$ or $J_3$ (more than two hundred times of $J_2$ and seventeen times of $J_3$), understanding the origination of $J_1$ is crucial to explain high Curie temperature of monolayer CrIAs. Here we come up with a direct exchange mechanism between Cr $d$ orbitals and I $p$ orbitals to interpreter both the sign and magnitude of $J_1$.

Two orbitals cannot be orthogonal if electrons intend to hop between them. In Cr-I-Cr path, they form a nearly right angular (Fig. 5a). Since $<d_{yz}|p_y> \neq 0$ and $<d_{yz}|p_z> \neq 0$, electrons hoping between $d_{yz}$ and $p_x$, $d_{yz}$ and $p_y$ are not forbidden. Because I $p$ orbitals have localized down spin moment, the electrons' spin of Cr atom must be up, according to Pauli Exclusion Principle. Therefore, I $p$ orbitals functionalized as a medium to induce FM coupling between Cr atoms, which caused the positive value, namely FM order, of

$J_1$.

To interpret the extremely large magnitude of nearest FM interaction $J_1$, we calculated the average energy level of selected orbitals as shown in Fig. 5b. The average energy level is calculated by

$$\langle E \rangle = \frac{\int_{-\infty}^{+\infty} g(E)EdE}{\int_{-\infty}^{+\infty} g(E)dE}$$

where $g(E)$ is density of states. The more close the energy level, the much strong the interaction. It is obvious that $d_{yz}$ orbital have almost same energy level as $p_y$ and $p_z$, thus induces the extremely strong interaction and provides an outcome which is the large value of exchange constant $J_1$. We noticed that $p_x$ also have similar energy level to $d_{yz}$, but the orthogonality $<d_{yz}|p_x> = 0$ principally forbid interaction within that two orbitals.

**Conclusion** In summary, we have proposed a new stable 2D CrIAs single crystal with room temperature magnetism. On the basis of first-principles calculation, we found that 2D CrIAs crystal is a FM semiconductor. The Curie temperature of FM CrIAs nanosheet is predicted to be 655 K based on the Heisenberg model (with magnetic anisotropic energy) Monte Carlo simulations. The high $T_c$ comes from direct exchange mechanism of Cr $d$ orbitals and I $p$ orbitals. The bandgaps of 2D CrIAs are 0.32 eV for majority spin and 3.31 eV for minority spin, both large enough to present semiconducting property at room-temperature. Our results point to a new intrinsic 2D semiconductor that combine ferromagnetism and thermal fluctuation resistance, providing the affirmative answer to one of the "125 questions" brought up by *Science*[34], and offering advantages over other 2D materials for room temperature applications in spintronic devices and quantum computation.

*Supporting Information*

The details of calculation method (Sec.1). The method of obtaining lattice parameters by optimizing them in confined dimensions and energies with respect to lattice parameters, atomic coordinates of FM CrIAs structure (Sec.2). The structure of different magnetic orientations and the derivation of magnetic interaction constants (Sec.3). Molecular dynamics results of energies, magnetic moment and structures with respect to time (Sec.4)

*Authors Information*

**F. Zhang**
 *Email*: zhang.fang@connect.um.edu.mo
 Orcid: 0000-0002-7313-0653
**X.Q Shi**
 Email: shixq@sustech.edu.cn
 Orcid: 0000-0003-2029-1506
**Z. K. Tang**
 Email: zktang@um.edu.mo
 Orcid: 0000-0001-9998-4940

*Corresponding authors*
**X.Q. Shi** & **Z. K. Tang**

*Acknowledgments*

This work in University of Macau is supported by the Science and Technology Development Fund (Grant No. FDCT-013/2017/AMJ), the Research & Development Grant for Chair Professor (Grant No. CPG-2019-00022-IAPME), the Multi-Year Research Grant (Grant No. MYRG-2018-00142-IAPME), the National Natural Science Foundation of China for Major Research Project (Energy-Oriented Photoelectric and Electrophotonic Materials, Grant No. 91733302). In Southern University of Science and Technology is supported by the Shenzhen Fundamental Research Foundation (Grant No. JCYJ20170817105007999), the Natural Science Foundation of Guangdong Province of China (Grant No. 2017A030310661). The computational resources are provided by the Center for Computational Science and Engineering of Southern University of Science and Technology.